\documentclass[a4paper,11pt]{article}
\usepackage{pos}

\title{Holographic description of the kaon gravitational form factor}

\author*[a]{Zhibo Liu}
\author[b]{Akira Watanabe}

\affiliation[a]{Department of Physics, Nagoya University,\\
  Nagoya 464-8602, Japan}

\affiliation[b]{National Institute of Technology, Oshima College,\\
Oshima 742-2193, Japan}

\emailAdd{liu\_zhibo@hken.phys.nagoya-u.ac.jp}
\emailAdd{watanabe.akira@oshima-k.ac.jp}

\abstract{
We compute the kaon gravitational form factor (GFF), using a bottom‐up holographic QCD approach that incorporates the $SU(3)$ flavor symmetry breaking through the strange quark mass.
We present the resulting $Q^2$ dependence of the kaon GFF and directly compare it with that of the pion.
In the high-energy limit, the obtained kaon GFF exhibits the $1/Q^2$  fall‐off, which agrees with the perturbative QCD prediction.
Furthermore, we extract the gravitational radius of the kaon and find it to be almost the same as that of the pion, with a slightly smaller value.
}

\FullConference{The 21st International Conference on Hadron Spectroscopy and Structure (HADRON2025)\\
 27 - 31 March, 2025\\
Osaka University, Japan\\}

\begin{document}
\maketitle

\section{Introduction}
The study of hadronic structure in the nonperturbative regime of quantum chromodynamics (QCD) remains a central challenge in modern hadron physics, while perturbative QCD successfully describes high-energy processes, studying static and dynamic properties, such as form factors, mass distributions, and energy-momentum tensor (EMT) matrix elements. In particular, the gravitational form factors (GFFs), defined via the hadronic EMT, encode fundamental information about the distribution of energy, pressure, and shear forces inside a hadron. Over the past decade, GFFs have attracted considerable attention because they provide unique insights into how QCD degrees of freedom (quarks and gluons) generate the mass and internal mechanical stability.  

Among various theoretical frameworks, bottom‐up holographic QCD (also known as AdS/QCD) has emerged as a powerful tool for modeling nonperturbative hadron structure. In this approach, one constructs a five‐dimensional effective action in the AdS background and introduces \(z\)-dependent fields to break the conformal invariance~\cite{Witten:1998qj, Karch:2006pv, Erlich:2005qh}. 
In this work, we focus on the kaon GFF within an \(N_f = 3\) bottom‐up holographic model that explicitly incorporates the strange quark mass and $SU(3)$ flavor symmetry breaking.  By computing the kaon EMT three‐point function in the holographic model, using the same parameter set previously fixed by fitting the lowest-lying baryon and meson spectra, we extract the dependence of the kaon GFF \(A_K(Q^2)\) on the momentum transfer \(Q^2\). In the high‐\(Q^2\) regime, our result exhibits the expected asymptotic behavior $A_K(Q^2) \sim 1/Q^2$, which is in agreement with the perturbative QCD prediction.
At low and intermediate \(Q^2\), we find that the kaon GFF closely parallels that of the pion, differing only slightly due to the strange quark mass. From the slope of \(A_K(Q^2)\) near \(Q^2 = 0\), we compute the kaon gravitational radius \(\langle r_G^2\rangle_K\), which is very close to, but marginally smaller than that of the pion, indicating a slightly more compact mass distribution attributable to the $SU(3)$ symmetry breaking.  

The remainder of this paper is organized as follows. In Sec.~\ref{section2}, we review the \(N_f = 3\) bottom‐up holographic QCD setup and present the calculation of the kaon EMT three‐point function, the extraction of \(A_K(Q^2)\), and numerical results for the GFF over a wide \(Q^2\) range. Finally, Sec.~\ref{conclusion} summarizes our findings and conclusions.

\section{Model setup and gravitational form factor}
\label{section2}
The 5D action for chiral gauge fields and the bifundamental scalar field \(X\) (dual to \(\bar q_R q_L\)) is ~\cite{Erlich:2005qh}
\begin{equation}
S_{5D} \;=\; -\int_{\varepsilon}^{z_0} d^5x\,\sqrt{-g}\,\rm{Tr}\Bigl[\,|D_M X|^2 - 3\,|X|^2 
-\frac{1}{4g_5^2}\bigl(F_L^{MN}F^{\,L}_{MN} + F_R^{MN}F^{\,R}_{MN}\bigr)\Bigr],
\label{eq:bulk_action}
\end{equation}
with 
\[
D_M X = \partial_M X + iL_M X - iX R_M,\quad
F_{L,R}^{MN} = \partial^M L^N - \partial^N L^M - i[L^M,L^N],
\]
and \(g_5^2 = 12\pi^2/N_c \).
We employ an \(N_c=3\) bottom‐up hard‐wall AdS/QCD model in the five‐dimensional AdS space:
\[
ds^2 \;=\; \frac{1}{z^2}\bigl(\eta_{\mu\nu}\,dx^\mu dx^\nu - dz^2\bigr),\qquad 0<z<z_0\sim1/\Lambda_{\rm QCD} = (322.5~\rm MeV)^{-1}.
\]
The UV boundary is \(z=\varepsilon\to0\), and the IR cutoff \(z_0\) implements the confinement.
The vacuum profile,
\begin{equation}
\begin{aligned}
\langle X\rangle &= X_0(z) = \frac{1}{2}\bigl(M_q\,z + \Sigma_q\,z^3\bigr),\\[6pt]
M_q &= \operatorname{diag}(m_q,m_q,m_s),\Sigma_q = \operatorname{diag}(\sigma_q,\sigma_q,\sigma_s),
\end{aligned}
\label{eq:X0}
\end{equation}
implements the explicit (\(m_s\neq m_q\)) and spontaneous chiral symmetry breaking.  The parameters \((m_q,\sigma_q)\) and \((m_s,\sigma_s)\) are fixed by \((m_\pi,f_\pi)\) and \((m_K,f_K)\)~\cite{Abidin:2009aj}.
 
Decomposing \(X=X_0\,e^{2i\pi^a t^a}\), the kaon wave function \(\phi_K(z)\) satisfies in the \(V_z=A_z=0\) gauge
\begin{equation}
\beta(z)\partial_z \pi(q,z) = q^2 \partial_z \phi(q,z), \qquad \partial_z\left(\frac{1}{z}\partial_z \phi(q,z)\right) = \frac{\beta(z)}{z}(\phi(q,z) - \pi(q,z)),
\label{eq:kaon_eom}
\end{equation}
where
\begin{equation}
\beta(z) = \frac{g_5^2 (v_s + v_q)^4}{16 z^2},\qquad v_{q(s)} = \frac{1}{2}(m_{q(s)}z + \sigma_{q(s)}z^3).
\end{equation}

\begin{figure}[tb]
\centering
\includegraphics[width=0.5\textwidth]{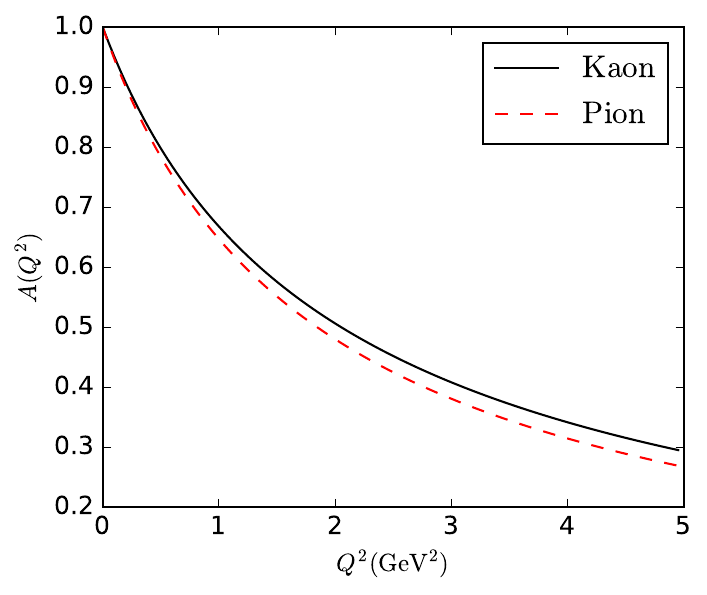}
\caption{
The gravitational form factors as a function of $Q^2$. The solid and dashed curves represent the kaon and pion results~\cite{Liu:2025vfe, Abidin:2008hn}, respectively.
}
\label{GFF}
\end{figure}

To obtain the kaon GFF \(A_K(Q^2)\), we introduce a transverse‐traceless graviton fluctuation \(h_{\mu\nu}(x,z)\) around the AdS metric:
\[
g_{MN} \to g_{MN} + h_{MN},\quad h_{zM}=0,\;\partial^\mu h_{\mu\nu}=0,\;h^\mu_{\ \mu}=0.
\]
The linearized coupling to the kaon arises from expanding the scalar kinetic term in Eq.~\eqref{eq:bulk_action} to first order in \(h\).  The 5D profile \(h(q,z)\) of the graviton, which is dual to the QCD EMT \(T^{\mu\nu}\), satisfies
\begin{equation}
\partial_z\Bigl(\frac{1}{z^3}\,\partial_z h(q,z)\Bigr) + \frac{q^2}{z^3}\,h(q,z) = 0,\qquad
h(q,\varepsilon)=1,\;\;\partial_z h(q,z_0)=0.
\label{eq:graviton_eom}
\end{equation}
Its normalized solution defines the bulk‐to‐boundary propagator for the EMT insertion with the momentum transfer \(q^\mu\), \(Q^2=-q^2\).
The matrix element of the QCD EMT between kaon states is parametrized by
\[
\langle K(p')\,|\,T^{\mu\nu}(0)\,|\,K(p)\rangle 
= 2P^\mu P^\nu\,A_K(Q^2) + \cdots,
\qquad P^\mu=\tfrac{1}{2}(p'+p)^\mu,\;\;Q^2=-(p'-p)^2,
\]
where only the form factor \(A_K(Q^2)\) is needed for the kaon.
In the holographic dual, we compute the GFF $A_K(Q^2)$:
\begin{equation}
A_K(Q^2) \;=\; \int_{0}^{z_0} \frac{dz}{z^3}\;h(q,z)\,\bigl|\phi_K(z)\bigr|^2.
\label{eq:AK_integral}
\end{equation}
Here \(\phi_K(z)\) solves Eq.~\eqref{eq:kaon_eom} and \(h(q,z)\) solves Eq.~\eqref{eq:graviton_eom}.  
The $Q^2$ dependence of $A_K$ and $A_\pi$ is shown in Fig.~\ref{GFF}, and it is found that $A_K$ decreases slightly slower than $A_\pi$ with $Q^2$.
By construction, \(A_K(0)=1\) (energy‐momentum conservation), and at large \(Q^2\),
\[
A_K(Q^2)= \frac{16 \pi^2 f_K^2}{Q^2},
\]
reproducing the perturbative QCD asymptotic scaling.
The gravitational radius of the kaon is extracted as
\[
\langle r_G^2\rangle_K = -6\frac{\partial A_K(Q^2)}{\partial Q^2}\bigg|_{Q^2 \to 0} = (0.35~\rm fm)^2.
\]

\section{Summary}
\label{conclusion}
We have investigated the GFF of the kaon within a holographic QCD framework employing a hard-wall cutoff, where the explicit introduction of the strange quark mass breaks the $SU(3)$ flavor symmetry. Beginning from the AdS/CFT correspondence, we established the essential mapping between 4D operators $\mathcal{O}(x)$ and 5D bulk fields $\phi(x,z)$, thus allowing for a holographic calculation of the kaon GFF via the EMT.

Our results for the kaon GFF demonstrate consistency with the perturbative QCD prediction at high momentum transfer $Q^2$. Additionally, we have computed the kaon gravitational radius and obtained the value $\langle r_G^2\rangle_K = (0.35~\rm fm)^2$, which is slightly smaller than that of the pion calculated within the similar holographic QCD approach.

Extending the present holographic QCD framework to study GFFs of mesons containing heavier quarks, such as the charm and bottom, would provide valuable insights into their internal structure and remain as an intriguing direction for future research.


\begin{thebibliography}{99}

\bibitem{Witten:1998qj}
E.~Witten,
Adv. Theor. Math. Phys. \textbf{2} (1998), 253-291
[arXiv:hep-th/9802150 [hep-th]].

\bibitem{Karch:2006pv}
A.~Karch, E.~Katz, D.~T.~Son and M.~A.~Stephanov,
Phys. Rev. D \textbf{74} (2006), 015005
[arXiv:hep-ph/0602229 [hep-ph]].

\bibitem{Erlich:2005qh}
J.~Erlich, E.~Katz, D.~T.~Son and M.~A.~Stephanov,
Phys. Rev. Lett. \textbf{95} (2005), 261602
[arXiv:hep-ph/0501128 [hep-ph]].

\bibitem{Abidin:2009aj}
Z.~Abidin and C.~E.~Carlson,
Phys. Rev. D \textbf{80} (2009), 115010
[arXiv:0908.2452 [hep-ph]].

\bibitem{Abidin:2008hn}
Z.~Abidin and C.~E.~Carlson,
Phys. Rev. D \textbf{77} (2008), 115021
[arXiv:0804.0214 [hep-ph]].

\bibitem{Liu:2025vfe}
Z.~Liu and A.~Watanabe,
[arXiv:2503.18747 [hep-ph]].

\end{thebibliography}
\end{document}